\documentclass[aps,prb,twocolumn,groupedaddress,showpacs,showkeys]{revtex4}
\usepackage{graphicx,amssymb,amsmath}

\begin{document}


\title{Pyrometric Measurement of the Temperature of Shocked Molybdenum}

\author{A. Seifter}
\email[Corresponding author e-mail:]{seif@lanl.gov}
\affiliation{Los Alamos National Laboratory\\Los Alamos, NM 87545}

\author{D. C. Swift}
\affiliation{Lawrence Livermore National Laboratory\\Livermore, CA 94550}

\date{\today}

\begin{abstract}

Measurements of the temperature of Mo shocked to $\sim$60 GPa and then released to $\sim$28 GPa
were previously attempted using high explosive driven flyer plates and pyrometry.  Analysis
of the radiance traces at different wavelengths indicates that the temporal evolution of the
radiance can be explained by a contribution from the LiF window to the measured thermal radiation.

Fitting the radiance traces with a simple model, supported by continuum dynamics studies which
were able to relate structures in the radiance history to hydrodynamic events in the experiment,
the contribution of the window was obtained and hence the temperature of the Mo sample.
The shock-and release temperature obtained in the Mo was 762$\pm$40K which is consistent with calculations
taking the contribution of plastic work to the heating into account. The radiance obtained for the LiF window
shows a non thermal distribution which can be described by a bulk temperature of 624$\pm$ 112K and hot spots
(less than 0.5\% in total volume) within the window at a temperature of about 2000K.

\end{abstract}



\maketitle

\section{\label{sec:level1}Introduction}

The measurement of temperature in shock physics experiments is of paramount importance since temperature is
required as constraint (additional to pressure and volume) for the development and credibility of robust multiphase
equation of states (EOS). The application of modern material strength models requires these EOS with accurate
temperatures and phase boundaries, particularly melt curves. Temperature is an important parameter in geophysics
and planetary sciences for an accurate description of planetary structure and astrophysical impacts.  Many measurements
of the EOS of planetary materials, and the pressure calibration on which diamond anvil cell studies rely, are made using shock
experiments, and rely on temperature corrections to infer states off the shock Hugoniot. General research and model development
for the response of materials to extreme dynamic conditions is focusing on multi-scale approaches, in which physics-based models
(as opposed to empirical relations) for thermally-activated processes including plastic flow, phase changes, and chemical reactions
requires an accurate knowledge of the temperature.  As a specific example, the melting curve and kinetics of Beryllium and Carbon
(Diamond) at elevated pressures are important to understand the behavior of ablator capsules for the national ignition facility (NIF)
to succeed in harvesting fusion energy from laser driven capsule implosions \cite{WSU-Workshop}.

Pyrometry is a form of thermal emission spectrometry in which the emission spectrum is collected in a small number of spectral bands
of significant bandwidth, compared with energy-dispersive spectrometers such as prisms and gratings.  Energy-dispersive spectrometers
are not practical for most shock experiments at temperatures below a few thousand Kelvin because the thermal emission is too weak to
take advantage of the relatively fine spectral resolution.  Pyrometry is thus the most promising technique applicable to many materials
and experimental configurations (besides Neutron Resonance Spectroscopy \cite{Yuan-NRS}$^,$\cite{Swift-NRS-Explanation} and Raman Spectroscopy
\cite{Gupta-Raman}, both with their advantages
and limitations) to achieve the goals stated above. Nevertheless, although pyrometry has been fielded on dynamic loading experiments
for more than four decades \cite{Kormer-Pyro1965} it still suffers from problems such as background light, thermal and non-thermal emission from the window
material and sample/window interface effects which are hard to take into account.

Neutron Resonance Spectroscopy (NRS) has been investigated as a possible bulk-temperature diagnostic for shock physics experiments,
capable of measuring the temperature within opaque samples. Initial experiments on shock loaded Mo showed significant discrepancy
with theoretical predictions. To investigate this discrepancy, microsecond-duration pyrometric measurements were performed on Mo
samples through LiF windows.  These experiments used a shock generation method that was the same as used in the NRS experiments,
and also included Doppler velocimetry measurements of the surface velocity history of the sample to verify the loading conditions.  Previous
analysis of the pyrometry experiments \cite{Seifter-Chamber8} did not succeed in extracting temperatures.  Here we report a more rigorous study of the effects of
shock and release waves on pyrometry measurements with a release window, identifying two-temperature population from pyrometry data.
We are able to correlate features of the radiance history at a level usually dismissed as too complicated for further analysis and provide
valuable insight into dynamic processes occurring in shock experiments using a window. This allows us to extract temperatures more
accurately and with greater confidence. The results presented here are the first simultaneous extraction of sample and window temperature
to our knowledge which enhances the ability to reliably interpret pyrometry data. This is a first step and key to future research and physics-based
understanding of response of condensed matter subject to dynamic loading and heating.
\section{Experimental Setup}
As in the NRS experiments, a shock was induced in the Mo sample by the impact of an Al disk accelerated to a speed of about 3.5 km/s by means of high
explosive (HE) gases \cite{Forest-Flyer}. A LiF window was glued to the backside of the Mo disk (where the measurement was made) in order to maintain high pressure
over an area accessible for contactless temperature measurements \cite{Asay-Book}. Emitted thermal radiation was focused onto a 1-mm core-diameter near-infrared (NIR)
fiber as well as onto a 1-mm core-diameter visible-glass, low OH fiber. Two different multi-channel pyrometers were used to infer thermal radiance over a
wide range of wavelengths: 5 channels in the visible and NIR wavelength region, using photomultiplier tubes (PMTs), and 4 channels in the NIR, using InSb
detectors. The velocity history at the Mo/LiF interface was measured by laser Doppler velocimetry, using a Velocity Interferometer System for Any Reflector
(VISAR) \cite{Barker-VISAR} in order to infer the pressure history applied to the sample. 

In the case of low-temperature pyrometry measurements, special precautions have to be taken in order to avoid background light, which can easily overwhelm
especially the short wavelength channels. These backgrounds can be generated either by bright HE gases blowing by the target, air lighting up due to shock
luminescence from nitrogen, ejecta or jets generated due to improper surface preparation, or sharp edges. In order to minimize the background, the experiments
were performed in a vacuum (10-3 Torr). Two CaF lenses were used to focus the thermally emitted light onto a 1-mm diameter NIR fiber \footnote{Chalcogenide glass fiber (C2)
from Amorphous Materials Inc., Garland, TX, www.amorphousmaterials.com}  (centered 2.5 mm off axis
at the fiber bundle) and a 1-mm diameter visible low OH fiber \footnote{Polymicro Technologies, Phoenix, AZ, www.polymicro.com} (centered 2.5 mm diametrically opposed).
The center fiber was a VISAR probe \footnote{One injection fiber in the center surrounded by seven receiving fibers each 100 $\mu$m in diameter} used to determine the
velocity of the sample after impact. The LiF window, 30 mm diameter by 20 mm thick, was included to maintain an elevated pressure in the Mo when the shock
reached its surface, avoiding a release to atmospheric pressure.  The LiF was attached to the Mo coupon using Loctite® 326 glue, which has been previously found \cite{Hereil-Loctite}
not to cause considerable amounts of thermal emission. This simplifies relating the surface temperature to the bulk temperature of the shocked sample.
However thermal light emission from both the glue layer and the LiF is a concern.

The NIR pyrometer collimates and divides the incoming light from a single NIR fiber into four spectral ranges, using three custom dichroic beamsplitters.
These four collimated beams are spectrally narrowed by bandpass filters centered at 1.8 $\mu$m, 2.3 $\mu$m, 3.5 $\mu$m and 4.8 $\mu$m and then refocused onto the 1 mm2
active area of the 50 MHz bandwidth InSb detectors using ZnSe lenses. The lower temperature limit of the two longer wavelength channels is about 340 K
(assuming an emittance of 1). For more information on the NIR instrument and on data analysis see Ref. \onlinecite{Seifter-Pyro-Alexandria}.

The 5 channel visible/NIR pyrometer also uses dichroic beamsplitters to spectrally divide the incoming light into five beams, which are then refocused onto
the active area of photo multiplier tubes (PMT). Also using bandpass filters, these five channels are centered at 505 nm, 725 nm, 850 nm, 1.23 $\mu$m and 1.59 $\mu$m.
A holographic notch filter is used to suppress the bright 532 nm laser light used for the VISAR measurements. More information on this instrument can be found in
Ref. \onlinecite{Holtkamp-VisPyrometer}.

\begin{figure}
\includegraphics*[width=1.5in]{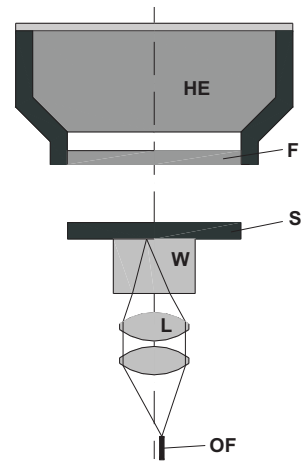}
\caption{Schematic setup of the experiment. S, molybdenum sample ($\emptyset$64 x 6 mm); W, LiF window ($\emptyset$25 x 20 mm), F; aluminum flyer ($\emptyset$64 x 5 mm,
20 mm gap between F and S); HE, 9501 high explosive (7 mm gap between HE and F); L, CaF$_{2}$ lenses ($\emptyset$25 mm, f=50 mm); OF, optical fibers ($\emptyset$1 mm C2
fiber, $\emptyset$1 mm low OH fiber \& VISAR fiber bundle).\label{fig:Fig1}}
\end{figure}

For further details on the HE driven flyer system see references \onlinecite{Forest-Flyer} and \onlinecite{Swift-Flyer}.
\section{Results}
Eight experiments have been performed at the ''Chamber-8'' high explosive experimental facility at the Los Alamos National Laboratory in September 2004; these experiments are
summarized in Table \ref{tab:Tab1}.

\begin{table}
\caption{Experiments performed. Surface condition (free surface or window), diagnostics (V, VISAR; IR, IR-pyrometer; vis, visible-pyrometer)\label{tab:Tab1}}
\begin{tabular}{|c|c|c|c|}
\hline
Exp $\#$&Surface&Diagnostics&Comments\tabularnewline
\hline
\hline
01&LiF window&V, IR&Data clipped\tabularnewline
\hline
02&LiF window&V, IR&Data lost\tabularnewline
\hline
03&LiF window&V, IR&Good data\tabularnewline
\hline
04&free surface&V, IR&HE problems, no data\tabularnewline
\hline
05&free surface&V, IR&Good data\tabularnewline
\hline
06&LiF window&V, IR, vis&Good data\tabularnewline
\hline
07&LiF window&V, IR, vis&Good data\tabularnewline
\hline
08&free surface&V, IR, vis&Good data\tabularnewline
\hline
\end{tabular}
\end{table}

\begin{figure}
\includegraphics*[width=3in]{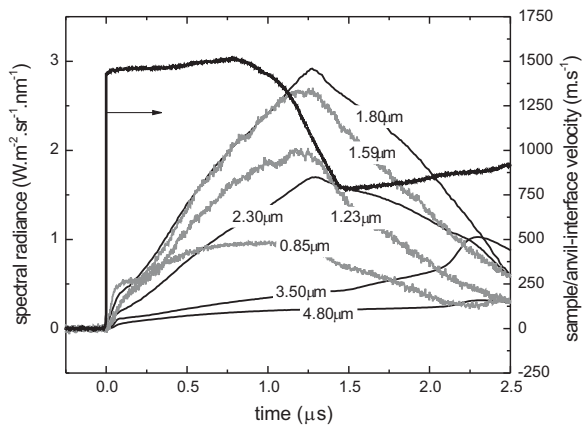}
\caption{Measured radiances for seven wavelengths and sample/window interface velocity for experiment $\#$06.\label{fig:Fig2}}
\end{figure}

Figure \ref{fig:Fig2} shows the radiance as function of time as well as the sample/window interface velocity for experiment $\#$06. The time of shock breakout is set to t = 0 $\mu$s.
If no unwanted background
light occurs the radiances are expected to be constant from the time of shock breakout until the release wave from the back of the flyer reaches the sample/window interface
(about 0.7 $\mu$s after breakout). At this time the sample/window interface is decelerating to a lower velocity as can be seen from the particle velocity trace.

Continuum dynamics simulations were used to investigate the origin of the varying signals in the shocked state by comparing them with the various hydrodynamic events
occurring in the experiments.  One and two dimensional (1D and 2D) simulations were performed, using general-purpose multi-physics hydrocodes.  Shock dynamics
calculations were also performed using the same material properties, solving the Rankine-Hugoniot equations to obtain shock states and hence wave speeds more precisely
than by derivation from the discretized hydrocode solutions \cite{Swift-Numerical-Solutions}.  The 1D and 2D hydrocodes both used a finite difference representation of spatial
fields, a second-order predictor-corrector algorithm for time integration, and artificial viscosity to stabilize shocks.  The 1D simulations, along the axis of symmetry, were
Lagrangian, avoiding inaccuracies caused by numerical advection, and used the LAGC1D hydrocode \cite{LAGC1D}.  The 2D simulations, axisymmetric in the axial-radial
plane, were Eulerian for robustness in treating highly distorted flow from sharp corners, with an operator-split third order van Leer flux limited method for advection, and
used the EUL2D hydrocode \cite{EUL2D}.

The simulations included the HE-driven flyer, the Mo target, and the LiF window.  The HE acceleration system itself was not modeled: the flyer was treated as flat and moving
at a constant speed, starting at the instant of impact.  In reality, the flyers were slightly dished, still accelerating slightly, and reverberating, but these details should not make
a significant difference to the pyrometry data.  The equations of state (EOS) used the cubic Grueneisen form with published parameters for each material \cite{Steinberg}. 
Some simulations were repeated using EOS from the SESAME tabular compendium \cite{Holian-LANL-EOS}; the difference was negligible. Strength was treated using the
Steinberg-Guinan model \cite{Steinberg-Guinan}, with published parameters for each material \cite{Steinberg}. In the 1D simulations, the cell size in the flyer, sample, and
window was 0.1, 0.05, and 0.2 mm respectively.  In the 2D simulations, the cell size was 0.1 mm.

\begin{figure}
\includegraphics*[width=3in]{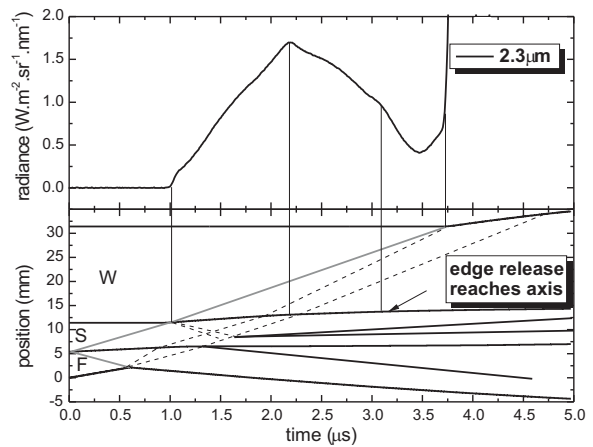}
\caption{x-t-diagram (lower part: black lines, boundaries between flyer (F), sample (S) and window (W); grey lines, shock front; dashed lines, head and tail of flyer release waves),
and measured spectral radiance at 2.3 $\mu$m for experiment 06 (upper part: the time has been shifted so that the breakout at the sample-window interface occurs at the same
time as at the simulated x-t-diagram).\label{fig:Fig3}}
\end{figure}

A position-time diagram for the key mechanical waves from the simulations is shown in figure \ref{fig:Fig3} (lower part) together with the measured spectral radiance at 2.3 $\mu$m
for experiment $\#$06.

The radiance increased from the time of shock breakout until about 2.1 $\mu$s. Then the radiance decreased at a rate slightly lower than the increase. The kink in the radiance at
about 3.1 $\mu$s occurs at about the time when the release from the edge of the window reaches the center of the sample-window interface, cooling it by adiabatic expansion.
The shock wave reaches the rear surface of the window at about 3.75 $\mu$s, at this time the signal of all wavelengths goes into saturation because of a bright flash that
occurs at this event.

The recorded pyrometer output signals for experiment $\#$01 have been clipped because the settings of the digitizer have been too sensitive (the increase of radiance with time
was not expected), the measured radiances for the other successful experiments look very similar to the ones shown in figure \ref{fig:Fig2} (for experiment $\#$06) and are not shown here.

The radiance traces for the free surface experiments are not analyzed and discussed here because, additionally to problems with background light, the surface temperature was
not homogeneous.  (This can be inferred from the response of different wavelengths shortly after the time of shock breakout. For more details on spatial temperature non-uniformities
see reference \onlinecite{Seifter-Pyro-2006}.) The background light at the free surface experiments was most likely caused by thermal radiation from APIEZON Q\copyright, a soft,
black, putty-like substance that was intended to absorb thermal radiation from hot jets formed at the edge of the sample/window interface.

\section{Data Analysis}

\begin{figure}
\includegraphics*[width=1.2in]{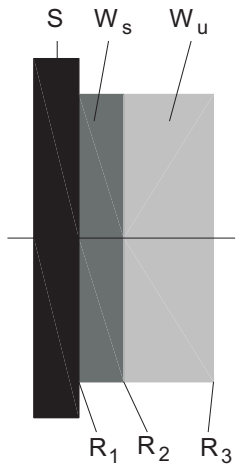}
\caption{Schematic of the window-sample combination. S, sample; W$_{s}$, shocked window; W$_{u}$, unshocked window, R$_{1-3}$, interface reflectivities.\label{fig:Fig4}}
\end{figure}

Figure \ref{fig:Fig4} shows the parts involved in the process of emission of thermal light for the sample/window combination used in these experiments, including the effect of the
interface between shocked and unshocked window material.

Thermal emitted radiance from the sample (L$_{sample}$) is multiply reflected between the interfaces R$_{1}$, R$_{2}$ and R$_{3}$. These reflectivities can be obtained from the
optical properties of the sample, the shocked and the unshocked window. The index of refraction of the shocked window can be estimated from the Gladstone-Dale relation 
\cite{Gladstone-Dale}, from this it can be seen that R$_{2}$ is very small and can be neglected\footnote{At a pressure 200kbar in the anvil the reflectivity R$_{2}$ is less than 0.1$\%$,
it increases to 0.25$\%$ at 500kbar and to 0.5$\%$ at 1Mbar.}. The index of refraction and extinction coefficient of the unshocked
window can be found in Ref. \onlinecite{Palik},
the absorption in this part can be neglected as well. The shocked part of the window can emit thermal radiation (L$_{window}$) as well as absorb thermal radiation emitted by
the sample. The emissive power of translucent materials (equivalent to emissivity of opaque materials) can be expressed by $(1-e^{-\alpha x})$ \cite{Gardon},
and the absorption can be expressed by $e^{-\alpha x}$ , where x is the thickness of the shocked layer and $\alpha$ is the absorption
coefficient. The absorption coefficient is related to the extinction coefficient by the following relation:
\begin{equation}
\alpha=\frac{4 \cdot \pi \cdot k}{\lambda_{0}}\label{eq:Equ1}
\end{equation}
where k is the extinction coefficient and $\lambda_{0}$ is the wavelength in vacuum. Thus the radiance from this assembly can be expressed as:
\begin{eqnarray}
L=\frac{(1-R_3)}{(1-R_1 \cdot R_3 \cdot e^{-2 \alpha x})} \cdot [L_{sample} \cdot \tilde{\epsilon} \cdot e^{- \alpha x}\nonumber \\ 
{} + L_{window} \cdot (1+R_1 \cdot e^{- \alpha x})  \cdot  (1-e^{- \alpha x})]\label{eq:Equ2} \end{eqnarray}
where $\tilde{\epsilon}$ is the effective emissivity of the sample into the window, taking the interface effects and absorption of the glue layer into account.
The emissivity is mainly determined by the optical properties (n and k) of the sample and the window, but can increase if the interface roughness increases
following shock breakout, for instance if the interface is Richtmyer-Meshkov unstable or because of flow or jetting from surface texture.

The first term in equation \ref{eq:Equ2} $(1-R_{3})/(1-R_{1} \cdot R_{3} \cdot e^{-2 \alpha x})$ is very close to 1
(between 0.98 and 1.0 in all practical cases) and can be set to 1 without compromising the accuracy of the analysis. The term
$(1+R_{1} \cdot e^{- \alpha x}) \cdot (1-e^{- \alpha x})$ can be simplified to 
$(1-e^{-2 \alpha x})$ without compromising the fitting results for the sample radiance, introducing an error in the window
radiation and absorption coefficient of less than 10\%, which is far below the uncertainty of the fitting parameters. Thus equation \ref{eq:Equ2} can be approximated as:
\begin{equation}
L=L_{sample} \cdot \tilde{\epsilon} \cdot e^{- \alpha x} + L_{window} \cdot (1-e^{-2 \alpha x}) \label{eq:Equ3} \end{equation}
The thickness x of the shocked layer is given by: $x=(u_{s}-u_{p}) \cdot t$ , where u$_{s}$ is the shock velocity in LiF (7.10 mm/$\mu$s at a pressure of 27GPa)
and u$_{p}$ is the particle velocity at the sample/window interface (1.45 mm/$\mu$s on release into LiF from a Hugoniot pressure of 64GPa in Molybdenum) and t is time after shock
breakout at the Mo/LiF interface. Equation \ref{eq:Equ3} was used to fit the radiance traces of experiments $\#$01, 03, 06 and 07. The results are shown in table  \ref{tab:Tab2}. 

\begin{table}
\caption{Fitting results for the experiments $\#$01, 03, 06 and 07. Where there are no data available (due to low signal level or poor quality of the fit) the cells have
been left blank.\label{tab:Tab2}}
\begin{tabular}{|c|c|c|c|c|}
\hline
Exp \# & $\lambda$ & $\epsilon \cdot L_{Mo}$ & $L_{LiF}$ & $a_{LiF}$ \tabularnewline
\hline
\hline
& ($\mu m$) & ($W/m^2 \cdot nm \cdot sr$) & ($W/m^2 \cdot nm \cdot sr$) & ($mm^{-1}$) \tabularnewline
\hline
&3.5&0.117&2.27&0.0109\tabularnewline
\cline{2-5}
\raisebox{1.5ex}{01}&4.8&0.051&0.229&0.0988\tabularnewline
\hline
&3.5&0.110& & \tabularnewline
\cline{2-5}
\raisebox{1.5ex}{03}&4.8&0.062& & \tabularnewline
\hline
&0.85& & &0.38694 \tabularnewline
\cline{2-5}
&1.23& &16.3&0.11307 \tabularnewline
\cline{2-5}
&1.59&&15.8&0.08272 \tabularnewline
\cline{2-5}
06&1.8&0.072&4.73273&0.0472 \tabularnewline
\cline{2-5}
&2.3&0.088&3.2168&0.0239 \tabularnewline
\cline{2-5}
&3.5&0.06401&0.85708&0.03602 \tabularnewline
\cline{2-5}
&4.8&0.02808&0.45&0.11792 \tabularnewline
\hline
&1.23&0.004&& \tabularnewline
\cline{2-5}
&1.59&0.025&& \tabularnewline
\cline{2-5}
&1.8&0.102&& \tabularnewline
\cline{2-5}
\raisebox{1.5ex}{07}&2.3&0.108&& \tabularnewline
\cline{2-5}
&3.5&0.132&& \tabularnewline
\cline{2-5}
&4.8&0.069&&0.1234 \tabularnewline
\hline
\end{tabular}
\end{table}

\begin{figure}
\includegraphics*[width=3in]{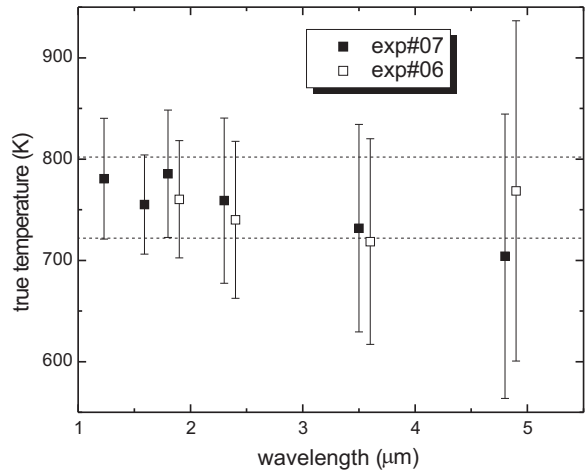}
\caption{Upper and lower temperature bounds as function of wavelength for experiments $\#$06 and 07 (the wavelengths for experiment $\#$06 have been shifted 0.1 $\mu$m to
the red for readability reasons).\label{fig:Fig5}}
\end{figure}

While the emissive power of the window material \cite{Gardone} was obtained from the fitting parameters, and hence the temperature of the shocked window can be obtained
directly, the inferred temperature of the sample depends strongly on the effective emissivity $\tilde{\epsilon}$ of the sample-window combination. Since this property is very hard to determine
experimentally, it is common practice to assume lower and upper emissivity bounds and use these bounds to calculate upper and lower bounds for the temperature
\cite{Vujnovic}. This has been done for experiments $\#$06 and 07 (where a sample radiance L$_{sample}$ was obtained) with the following lower $\epsilon_{l}$ and upper
$\epsilon _{u}$ emissivity
bounds: $\lambda$=1.23 $\mu$m: $\epsilon_{l}$ = 0.2, $\epsilon _{u}$ = 1.0; $\lambda$=1.59 $\mu$m: $\epsilon_{l}$ = 0.15, $\epsilon _{u}$ = 0.8; $\lambda$=1.80 $\mu$m:
$\epsilon_{l}$ = 0.12, $\epsilon _{u}$ = 0.8; $\lambda$=2.30 $\mu$m: $\epsilon_{l}$ = 0.1, $\epsilon _{u}$ = 0.6; $\lambda$=3.50 $\mu$m: $\epsilon_{l}$ = 0.08,
$\epsilon _{u}$ = 0.4; $\lambda$=4.80 $\mu$m: $\epsilon_{l}$ = 0.05,$\epsilon _{u}$= 0.3. These estimates are based on the room temperature values (obtained from the optical properties)
for polished surfaces \footnote{The normal spectral emissivity for a flat surface is obtained by: $\epsilon _{\perp}=\frac{4 \cdot n_{1} \cdot n_{2}}{(n_{1} + n_{2})^2 + k_{2}^2}$
with n$_{1}$ and n$_{2}$ index of refraction of the anvil and the sample and k$_{2}$ the extinction coefficient of the sample}. \cite{Palik} and the melting temperature
values at ambient pressures \cite{Vujnovic}. The calculated true temperature bounds for experiment $\#$06 and 07 as function of wavelength are shown in
figure \ref{fig:Fig5} (for readability reasons the wavelengths for experiment $\#$06 have been shifted by 0.1 $\mu$m to the red).

It can be seen that the uncertainty due to unknown emissivity is far higher for longer wavelengths, as expected.  (For an explanation see reference \onlinecite{Seifter-Pyro-2006}.)
The temperature ranges for all wavelengths overlap between 722 K and 802 K; this is the range where the temperature of the sample shocked to a Hugoniot pressure of 63.9$\pm$2.4 GPa
and released into LiF (at an interface pressure of 27.1$\pm$1.0 GPa) is determined. The Hugoniot pressures in the sample have been obtained from the measured particle
velocity at the sample window interface, the pressures in the LiF window have been obtained from jump conditions and known shock properties of the window material,
through the published EOS and strength model \cite{Steinberg}.

\begin{figure}
\includegraphics*[width=3in]{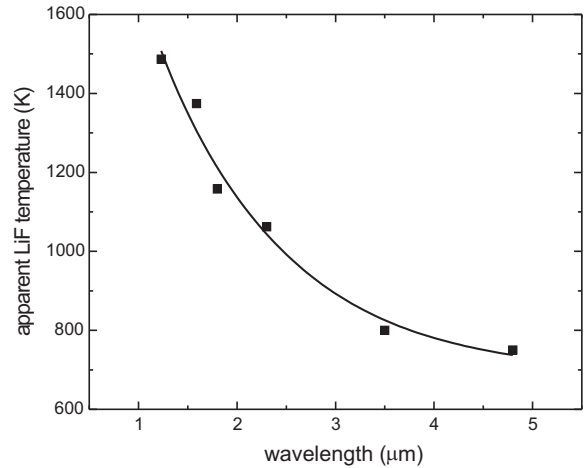}
\caption{Apparent window temperature as function of wavelength (squares) from experiment $\#$06. The solid line is the result from fitting the measured data with a simple two
temperature model \cite{Seifter-Pyro-2006}, the results of the fitting process are presented in the text below.\label{fig:Fig6}}
\end{figure}

The temperature of the LiF window as function of wavelength from experiment $\#$06 is shown in figure \ref{fig:Fig6}.

\begin{figure}
\includegraphics*[width=3in]{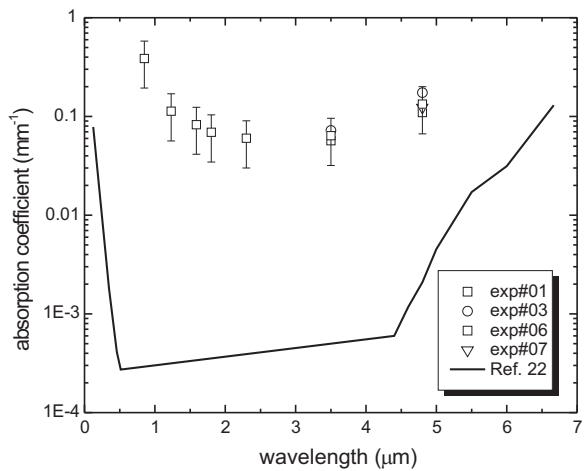}
\caption{Absorption coefficient for shocked LiF (27.1$\pm$1.0 GPa) as function of wavelength for all wavelengths and experiments where these data could be obtained from the fitting
process. The solid line shows literature values for LiF at standard conditions.\label{fig:Fig7}}
\end{figure}

If the window is at a homogenous temperature and no background light occurs, the measured window temperature should be independent of wavelength. The increase
of the measured window temperature (apparent temperature) with decreasing wavelength indicates a non-homogenous window temperature. Fitting the results with a simple
two temperature model (for details see reference \onlinecite{Seifter-Pyro-2006}) gives a temperature of the window of 624$\pm$112 K with hot cells at a temperature of 1945$\pm$210 K.
The area fraction (as seen from the pyrometer) of these hot cells was determined to be 3.0+1.5$\%$, corresponding to a volume fraction of less than 0.5$\%$. The absorption
coefficient as function of wavelength for the shocked LiF window is shown in figure \ref{fig:Fig7}. The solid line shows data from reference \onlinecite{Palik} at room temperature and
ambient pressure. It can be seen that the absorption coefficient increases by roughly two orders of magnitude compared to the literature data independent of wavelength.

It can be seen from figure \ref{fig:Fig3} that the radiance increases as long as the thickness of the shocked layer in LiF is increasing (from shock breakout at the Mo-LiF interface at
1.0 $\mu$s until the release wave from the back of the aluminum flyer reaches this interface at about 2.1 $\mu$s). The thickness of the shocked LiF layer decreases after t=2.1 $\mu$s
because the release wave propagates at a higher velocity than the shock wave and hence the measured radiation decreases between t=2.1 $\mu$s to about 3.7 $\mu$s. At 3.7 $\mu$s
the shock front in the LiF reaches the free surface of the window, and the measured radiance increases abruptly, which is assumed to be due to non thermal light emission when
the window fractures on release from the shocked state and possibly the air shock.

\begin{figure}
\includegraphics*[width=3in]{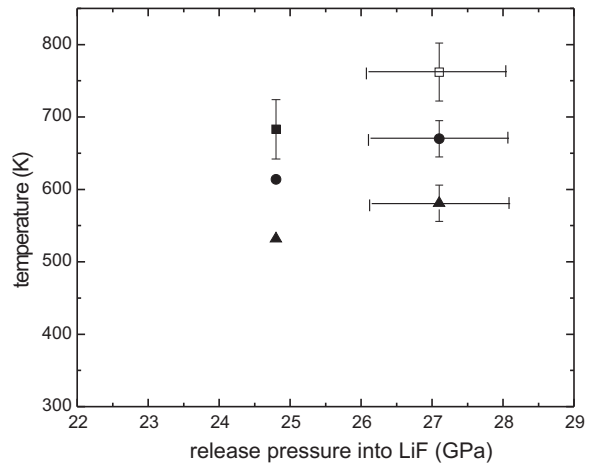}
\caption{Measured and calculated temperatures for molybdenum releasing into LiF for two different release pressures. Open square, this experiment; full square, gas gun
experiment [\onlinecite{Swift-Molybdenum}]; circles, calculations taking strength into account via the Steinberg-Guinan model [\onlinecite{Steinberg-Guinan}]; triangles,
calculations ignoring strength effects [\onlinecite{Steinberg-Guinan}].\label{fig:Fig8}}
\end{figure}

From fitting the spectral radiance traces at different wavelengths and assuming lower and upper bounds for the effective emissivity of the sample/window assembly, a sample
temperature (shocked to 63.9$\pm$2.4 GPa and released to a pressure of 27.1$\pm$1.0 GPa into LiF) at the interface of 762$\pm$40 K could be obtained. The release temperature
into LiF was calculated to be 670$\pm$25 K using the Steinberg-Guinan strength model to predict the contribution to heating from plastic work [\onlinecite{Swift-Molybdenum}]. The resulting
temperatures are in relatively good agreement \footnote{For both calculations (with and without strength) the temperature at 24.8 GPa is about 55 K lower
than the one at 27.1 GPa. The measured data point at 24.8 GPa is 79$\pm$35 K lower than the 27.1 GPa one.} with experiments performed on a gas gun at a slightly
lower pressure (T=683$\pm$41 K at a Hugoniot pressure of 58.7 GPa, with a release pressure of 24.8 GPa into LiF). A summary of all experimental and calculated results
can be seen in figure \ref{fig:Fig8}.

The measured temperature in the LiF window is 624$\pm$112 K with hot spots (3.0$\pm$1.5$\%$ in area) with a temperature of 1945$\pm$210 K. These hot spots are believed to occur
due to the slightly non uniaxial loading of the sample and window [\onlinecite{Swift-NRS-Explanation}] due to the dishing of the flyer plate which most likely causes local heating
by shear strain localization. This localized heating in LiF windows was also observed in infrared imaging experiments [\onlinecite{Wilke}]. The temperature of the LiF window was
calculated to be 569$\pm$15 K [\onlinecite{Swift-Molybdenum}] which is in excellent agreement with the temperature inferred from the experiment.
An analysis of the free surface experiments was not possible because of temperature non uniformities combined with background light caused by the hot APIEZON Q\copyright 
 (used to block thermal emission from hot jets from the edges, as discussed above).

\section{Conclusions}

Reliable pyrometric temperatures were obtained at shocked and released surfaces despite an \it o\rm(100$\%$) background from shocked window material. 
This improvement was achieved by using the emission spectrum to infer a volume fraction of hotspots in the window, and was supported by relating changes in the emitted
radiance to predictions of hydrodynamic events in the window. The measured temperature for Mo at a Hugoniot pressure of 63.9$\pm$2.4 GPa released into LiF to an
interface pressure of 27.1$\pm$1.0 GPa of 762$\pm$40 K is in good agreement with temperatures measured using a powder gun at a slightly lower pressure
(58.7 GPa released to 24.8 GPa) of 683$\pm$41 K. The measured temperatures are slightly higher than calculations using the Steinberg-Guinan model [\onlinecite{Steinberg-Guinan}]
to take strength effects into account. It is not clear whether the remaining discrepancy is caused by additional sources of thermal emission not taken into account, or inaccuracy
in the models used in the simulations.  However, this level of agreement is unusually good for thermal emission from shocked metals.

Besides the release temperature of Mo, the shock temperature of LiF at a pressure of 27.1$\pm$1.0 GPa was also obtained and is in good agreement with calculations. 
The shock temperature in the window was much less sensitive to the strength model.

Because of problems controlling the thermal background, free surface temperatures could not be extracted from the experiments without a window.

%


\begin{acknowledgments}
The work was performed under the auspices of the U.S. Department of Energy under contracts W-7405-ENG-36 and DE-AC52-06NA25396.
\end{acknowledgments}

\bibliographystyle{apsrev}
\bibliography{RefsAchim}

\end{document}